\documentclass{PoS}
\usepackage{amsmath}

\title{The K/$\pi$ ratio and the lifespan of the fireball}

\ShortTitle{The K/$\pi$ ratio and the lifespan of the fireball}

\author{\speaker{Boris Tom\'a\v{s}ik}\\
        Fakulta pr\'irodn\'ych vied, Univerzita Mateja Bela,
	Tajovsk\'eho 40,\newline 97401 Bansk\'a Bystrica, Slovakia\\
	and \'Ustav vedy a v\'yskumu, Univerzita Mateja Bela,
	Cesta na Amfite\'ater 1,\newline 97401 Bansk\'a Bystrica, Slovakia\\
        E-mail: \email{boris.tomasik@umb.sk}}

\author{Evgeni E Kolomeitsev\\
        University of Minnesota, School of Physics and Astronomy,
	116 Church Street SE, Minneappolis, 55455 Minnesota, USA\\
        E-mail: \email{kolomeitsev@physics.umn.edu}}

\abstract{%
We construct a hadronic kinetic model for description of
excitation functions of multiplicity ratios $K^+/\pi^+$, $K^-/\pi^-$,
and $\Lambda/\pi$. It is shown that the model is able to describe
data rather well under the assumption that the total lifespan
of the fireball is decreasing function of the collision energy
for SPS energies above 30~$A$GeV. Thus in order to identify the
onset of deconfinement the proposed model will have to be checked
against other kinds of data until hadronic description  can be safely
falsified.
}

\FullConference{Critical Point and Onset of Deconfinement\\
                 July 3-6 2006\\
                 Florence, Italy}

\begin{document}

\section{Motivation: ``the horn''}
\label{motiv}

Very often at this conference possible onset of deconfinement in
Pb+Pb collisions at $\sqrt{s_{NN}} \approx 8\,\mbox{GeV}/c$ has
been discussed. Indeed, there are (at least) three interesting
excitation functions showing non-monotonic structures just in this
collision energy region. Insiders describe them as ``the horn'',
``the step'', and ``the kink'' (for review see e.g.\ \cite{roland}).
The first relates to the ratio of
multiplicities of positively charged kaons and pions. The
excitation function of this quantity first rises rapidly to reach
a peak  around the quoted energy, and then drops down rather sharp
and levels off around $\sqrt{s_{NN}}\approx 9\, \mbox{GeV}/c$
(see e.g.\ \cite{lung}). The
second is the mean transverse momentum of kaon spectra which is
rising at low energies, is flat roughly in the same region where
we observe the horn, and then rises again. Finally, the kink
describes the excitation function of the ratio of produced pions
per participating nucleon: it starts growing more rapidly just at
the same collision energy as the peak of the horn.

Some authors speculate that these observations indicate the onset
of deconfined matter. Statistical Model of Early Phase (SMES)
\cite{smes} reproduces data while it makes a clear assumption that
the deconfined phase is reached in the early phase of collision.
It is rather schematic, though, and the assumption of almost
immediate chemical equilibrium could appear surprising. A kinetic
calculation reproducing  data on $K/\pi$ ratio has also been
performed \cite{kincal}. It does not use the argument of early
equilibrium. These are positive indications: models which do
assume the onset of deconfinement and are capable to accomodate
data.

The evidence for transition to plasma of colour charges, however,
must be based on exclusion of any hadronic interpretation of data.
Several studies went in this direction. Transport generators RQMD
and HSD \cite{brat} overpredict the $\pi^+$ multiplicity. Another
BUU model \cite{wagner} does not show any peak in the $K/\pi$
ratio. A three-fluid hydrodynamic model \cite{toneev} with
underlying hadronic equation of state reproduces multiplicities of
$K^+$ and pions, but over-predicts the multiplicity of $K^-$
(which is puzzling since the total strangeness must vanish in both
theory and experiment). Statistical model assuming full chemical
equilibrium \cite{statmodel} shows a maximum of the $K^+/\pi^+$
ratio of multiplicities, which is not as high and sharp as the
observed one, though. Better agreement with data is obtained if
strangeness suppression factor is included in the fit
\cite{becat}. This opens the question which values of suppression
factor are reached during the fireball evolution.

From all these studies a conclusion seems to be emerging that
deconfinement must set in at lowest SPS energies, for otherwise
data cannot be accommodated. We shall play {\em advocatus diaboli}
here and test a hadronic model on the description of the horn. The
horn shall be reproduced successfully, which means that our model
will have to be proved on other data also before we can safely
exclude it or confirm its validity.

\section{Strangeness production in nuclear collisions}
\label{stranprod}

Strangeness has been proposed many years ago as a good signal of
deconfinement \cite{rafel}. The simplest argument was that it is
energetically cheaper to produce a pair of $s\bar s$ quarks in
deconfined medium than to produce a pair of $S=\pm 1$ hadrons in
hadronic gas. The former requires energy around 300~MeV while
most favorable hadronic reaction is $\pi+N\to K+\Lambda$ with
threshold of 530~MeV above  the incoming masses. Thus rates for
production of strangeness are naturally expected to be higher
in plasma and this should lead to larger total production.

In general, there are two handles to control the final produced
amount of strange particles: energy available for strangeness
production and time. The former influences the rates. More
available energy opens phase space for strangeness-producing
reactions and thus increases production rates. The role of time is
also rather clear. In a situation of strangeness
under-representation the time evolution of the system is always
directed towards increasing strangeness content. The longer the
system lives, the closer it comes to equilibrium.

\begin{figure}[ht]
\centerline{\includegraphics[width=.8\textwidth]{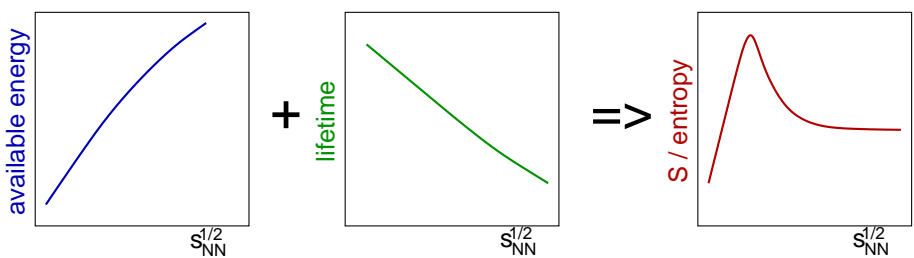}}
\caption{
We assume that energy available for production of strange particles
increases with collision energy while the lifetime of fireball decreases
with increasing collision energy. Then, these two excitation functions might
combine into a peak in strangeness-to-entropy ratio as a function of
collision energy.}
\label{f:cartoon}
\end{figure}

The $K^+/\pi^+$ ratio is basically a measure of strangeness-to-entropy
ratio. Now we wonder if it is possible to combine the available
energy and total lifespan of the fireball as functions of collision
energy in such a way that we reproduce the horn in strangeness-to-entropy?
It is natural to expect that the available energy will be an increasing
function of the collision energy. Next, we shall make the assumption
that the total lifespan of the fireball shall {\em decrease} as collision
energy increases (Fig.~\ref{f:cartoon}).
When combining these two features it is imaginable that the
strangeness-to-entropy ratio would show a peak. The initial rise is
due to increasing available energy which opens  larger phase space for
production of strange particles. The decrease following
the peak is due to shortening time available for strangeness production.

We shall construct  a kinetic model for strangeness
production and test this hypothesis.  In calculating the
rates we shall only assume hadronic degrees of freedom.

\begin{figure}[t]
\centerline{\includegraphics[width=.4\textwidth]{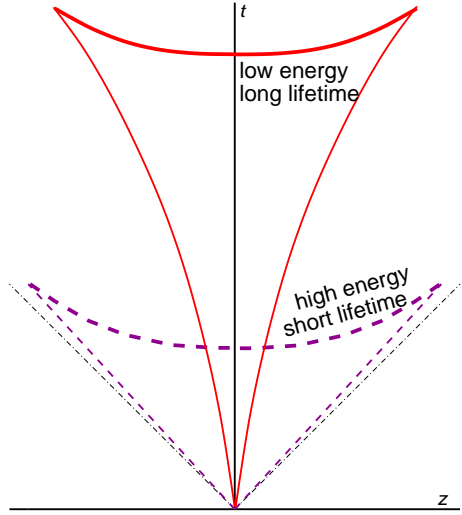}}
\caption{
Longer lifetime of a fireball
does not necessarily imply larger extent in longitudinal direction.
Evolution scenarios and freeze-out hypersurfaces in a space-time
diagram for fireballs created at lower collision energy with strong
stopping and re-expansion (solid line), and higher collision energy in
nuclear transparency regime (dashed line).}
\label{f:lspan}
\end{figure}

Before proceeding to the explanation of the model let us comment
on the assumption that the lifespan of the fireball decreases with
increasing collision energy. There is a widely spread lore that
HBT data determine the lifetime to be around 10~fm/$c$. This is
not true! It is important to realize that the measurement of total
lifetime via femtoscopy is model-dependent. What is measured is
the longitudinal length of homogeneity region. This length can
be translated into total lifespan only with help of a model
(Fig.~\ref{f:lspan}). Usually, Bjorken model \cite{bjork} is used
and this is where the quoted 10~fm/$c$ comes from. If, however,
there is nuclear stopping and the expansion of the fireball is to
a large extent built up from pressure, then Bjorken scenario is
invalid and the measured longitudinal size may be reached after a
longer lifespan. An implication of this is that longer lifespan
does not necessarily mean larger freeze-out volume and larger
multiplicity.
Various physics can lead to prolonged lifetime. Stopping and reexpansion
is one possibility. Another possibility is softening of the equation
of state, i.e., weak pressure gradients which do not enough accelerate
expansion.


\section{A kinetic model of strangeness production}
\label{model}

We shall perform numerical kinetic calculation of strangeness
production. In its spirit, the model is very similar to early
works on strangeness production \cite{kora,komura,kame} and later
flavour-kinetic \cite{knoll,knoll2} or hydro-kinetic \cite{brko}
simulations. In contrast to the later, we do {\em not} calculate
the space-time evolution in framework of a hydrodynamic model. We
{\em parametrise} it. The reason for this approach is that we know
from femtoscopic studies that hydrodynamic simulations do not
reproduce sizes of the fireball at breakup. Therefore, there is no
reason to pretend that they give the correct space-time evolution
of the fireball. On the other hand, if we {\em parametrise} the
evolution, we have the freedom to explore various different
fireball evolution scenarios and identify those which best correspond
to data. We shall go this way.

Since we shall only calculate the ratios of multiplicities and
not the multiplicities themselves we do not need to calculate the
volume of the system; it shall suffice to know the {\em average} densities
of species. In particular, we shall calculate and evolve
densities of kaons with $S=1$. These follow the master equation
\begin{equation}
\frac{d \rho_K}{d\tau} = \frac{d}{d\tau}\, \frac{N_K}{V} =
-\frac{N_K}{V} \frac{1}{V} \frac{dV}{d\tau} + \frac{1}{V}
\frac{dN_K}{d\tau}\, .
\end{equation}
Obviously, the first term on the right hand side includes the expansion
rate $1/V\, dV/d\tau$ and stands for the change of density solely due
to expansion. The second term represents the change of density resulting
from kaon production and/or annihilation. It can be split into two
terms: production rate and annihilation rate.  Thus the resulting
master equation generally reads
\begin{equation}
\label{maseq} \frac{d \rho_K}{d\tau} = \rho_K \, \left ( -
\frac{1}{V}\, \frac{dV}{d\tau}\right ) + {\cal R}_{\rm gain} -
{\cal R}_{\rm loss}\, .
\end{equation}

In hydro-kinetic or flavour-kinetic approaches, the expansion rate
follows from hydrodynamic calculation. We shall use an {\em
ansatz} for the evolution of energy and baryon densities. The
ansatz concerning baryon density evolution will enter into our
calculation through the expansion rate.

\subsection{Production and annihilation}
\label{proa}

From Eq.~\eqref{maseq} we calculate densities of $K^+$, $K^0$,
$K^{*+}$, and $K^{*0}$. Throughout our calculation we assume that all
particles keep their vacuum properties.

The gain term in Eq.~\eqref{maseq} includes two
types of contributions
\begin{equation}
{\cal R}_{\rm gain} = \sum_{i\, j\, X} \langle
v_{ij}\sigma_{ij}^{KX}\rangle \frac{\rho_i \rho_j}{1+\delta_{ij}}
+ \rho_{K^*} \Gamma_{K^*}\, ,
\end{equation}
where the sum goes over two-to-N processes leading to production
of kaons and the second term is for $K^*$ decay into $K$.
In angular brackets we have the cross-section multiplied with relative
velocity of the relative pair and averaged over all relative velocities.
In this averaging we assume thermal distribution of velocities. Thus kaons
are assumed to be in {\em thermal} equilibrium until the decoupling. The loss
term includes processes which destroy kaons and is obtained in a
similar way
\begin{equation}
{\cal R}_{\rm loss} = \sum_{i\, X} \langle v_{Ki}
\sigma_{Ki}^X\rangle \frac{\rho_K \rho_i}{1+\delta_{Ki}}\, .
\end{equation}

In practice,  it is impossible to include {\em all} reactions which create
or destroy a kaon. Most  of them, particularly those involving heavier
particles, have unknown and/or poorly constrained cross-sections. Their
influence, however, is not so important. Due to large mass the abundance
of heavy particles is small and the rate of corresponding  processes is
low. We implement following reactions in our calculation:
\begin{itemize}
\item Associated production of kaon and hyperon in reactions of
pions with $N$ and/or $\Delta$. Kaon annihilation on hyperons
leading to $N$ or $\Delta$ with pion is also taken into account.
\item Meson reactions of $\pi\pi$, $\pi\rho$, and $\rho\rho$ leading to
$K\bar K$ production are included in both directions: creating and
destroying kaons.
\item Production of $K^*$ from $K$ and $\pi$ and its decay.
\item Reactions of $\pi Y \leftrightarrow K\Xi$, these are also
included in both directions.
\item Baryon-baryon reactions of $NN$, $N\Delta$, and $\Delta\Delta$ which
produce kaons lead to more than two particles in the final state
and are included only in kaon gain term.
\end{itemize}

Species other than kaons with $S=1$ are treated according to
equilibrium assumptions. The non-strange species are assumed to be
in chemical equilibrium. This follows from the assumption that
inelastic processes among them are quick and able to keep the
chemical equilibrium. Species with $S<0$ (i.e.\ those containing
strange quark) must balance the abundance of strange antiquarks
contained in kaons. We assume, that strong interactions which do
not create but just reshuffle strange quarks between two particles
are fast enough to keep the $S<0$ particles in {\em relative}
chemical equilibrium.

In principle, {\em all} reactions can be treated kinetically.
However, it appears that the assumptions of equilibria bring the
model closer to how  real process actually runs. There are many
reactions with poorly known or unknown cross-sections. Unlike in
strangeness production, here these processes may have important
influence on the relaxation time. Due to the large variety of
possible processes and presumably large cross-sections the
relevant relaxation times shall be small enough in order to keep
the non-strange sector of the whole system in equilibrium.
Incomplete calculation with wrong cross-sections, on the other
hand, may deviate from this strongly. Since a complete kinetic
calculation is technically not possible, it is safer to {\em
assume} (partial) chemical  equilibrium and calculate abundances
from this assumption.

Finally, let us note that we included  no antibaryons into our
calculations. Since the system represents  baryon-rich environment, we
do not expect a big error connected with this simplification. The amount
of antibaryons grows with collision energy. At the highest SPS energy the
ratio of $\bar \Lambda/\Lambda$ multiplicities is around 10\%, so this
gives us an upper estimate for the error we commit here.

\subsection{The ansatz for expansion}
\label{exp}

It will be assumed that the evolution of fireball densities
consists basically of two periods. First, the expansion is accelerated
due to internal pressure of the matter. Here we shall assume the simplest
parametrisation of accelerating expansion: quadratic dependence
on time. Afterwards, the expansion
turns into a scaling scenario with power law decrease of
baryon density, as it is commonly assumed and in accord with femtoscopy
measurements.  Since baryon number is conserved quantum charge,
the time dependence of average baryon density gives the evolution
of inverse volume. We assume that is goes like
\begin{equation}
\label{npar} \rho_Q(\tau) = \left \{
\begin{array}{lcl}
\rho_{Q0} (1 - a\tau - b\tau^2)^\delta & : & \tau < \tau_s \\
\frac{\rho_{Q0}^\prime}{(\tau - \tau_0)^\alpha} & : & \tau \ge
\tau_s
\end{array} \right . \, .
\end{equation}
where $\rho_{Q0}$, $a$, $b$, $\rho_{Q0}^\prime$, $\tau_0$,
$\tau_s$, $\alpha$ and $\delta$ are parameters which can be tuned.
They can be more conveniently expressed in terms of total
lifespan, initial baryon density, initial expansion rate, maximum
expansion rate, etc. \cite{ktopi}. We shall change these
parameters and thus explore a wide range of possible evolution
scenarios. Our models may vary ``in between'' the commonly known
Bjorken \cite{bjork} and Landau \cite{dau} models.

The energy density is parametrised similarly to baryon density.
The equation of state enters in a  simple way by relating
the equation for energy density to that for baryon density through
the exponent $\delta$
\begin{equation}
\label{epar}
\varepsilon(\tau) = \left \{
\begin{array}{lcl}
\varepsilon_{0} (1 - a\tau - b\tau^2) & : & \tau < \tau_s
\\ \frac{\varepsilon_{0}^\prime}{(\tau - \tau_0)^{\alpha/\delta}} & : &
\tau \ge\tau_s
\end{array} \right . \, .
\end{equation}

\subsection{Initial and final conditions}

We already specified the master equation and its ingredients. It
remains to determine the initial conditions. Clearly, strange
particles are first produced in  primordial collisions of incoming
nucleons. The coresponding ``initial'' densities of $K^+$ and
$K^0$ are estimated from data on proton-proton collisions and
extrapolations to nucleon-nucleon collisions \cite{garo}. The
initial densities of species with $S<0$ are then obtained by
making use of the requirement that total strangeness must vanish.
Ratios of these densities are fixed according to the
assumption of relative equilibrium.

In order to choose a specific model, we must fix  parameters
which appear in Eq.~\eqref{npar}. The most important quantities which
are selected are the initial energy density $\varepsilon_0$ and the
total lifespan $\tau_T$. Furthermore, our simulation must arrive
to the final state which corresponds to  chemical freeze-out.
Thus we fix the final values of energy density, baryon density,
and the density of third component of isospin. These we infer from
chemical freeze-out fits of \cite{becat}.

In this way,
we are guaranteed by construction to arrive at the correct
final state energy and baryon density. Notice, however, that this
does not automatically warrant the correct temperature. At fixed energy
density the temperature depends on the effective number of degrees of
freedom and thus the resulting temperature will depend on the
amount of produced strange particles, which is calculated kinetically.

The argument can be pushed further. {\em If} the density of
strange particles comes out correct, then the temperature and
baryochemical potential are also correct. In such a case,
abundances of {\em  all} species are correct, since they are
calculated statistically.

\subsection{Examples: time evolution of densities}

Before presenting the results, let us illustrate the time evolution
of densities.
\begin{figure}[t]
\centerline{\includegraphics[width=.5\textwidth]{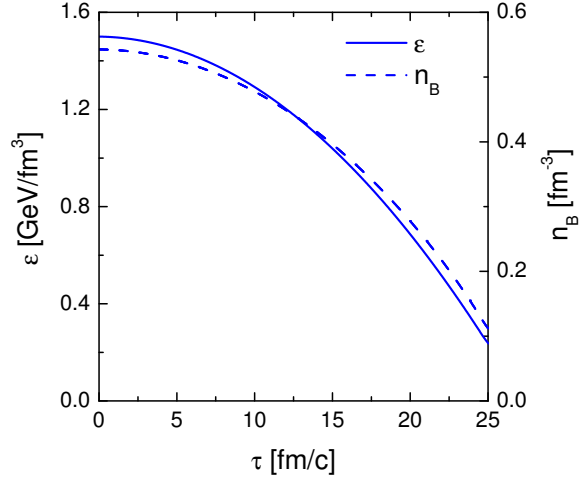}}
\caption{Time evolution of energy density (solid line, left scale)
and baryon density (dashed line, right scale) in a model which
reproduces the data  from Pb+Pb collisions at projectile energy
30~$A$GeV.}
\label{f:eden}
\end{figure}
We see in Fig.~\ref{f:eden} that most of the time the fireball
stays in the accelerating phase characterised by the quadratic
dependence of density on time. This feature is rather general in
our model. It is so because of the requirement of rather low
initial energy density. If the power-law time dependence of the
second phase was realised for a longer time, this would require
very high initial energy density reaching up to tens of
GeV/fm$^3$.

\begin{figure}[t]
\centerline{%
\includegraphics[width=.3\textwidth]{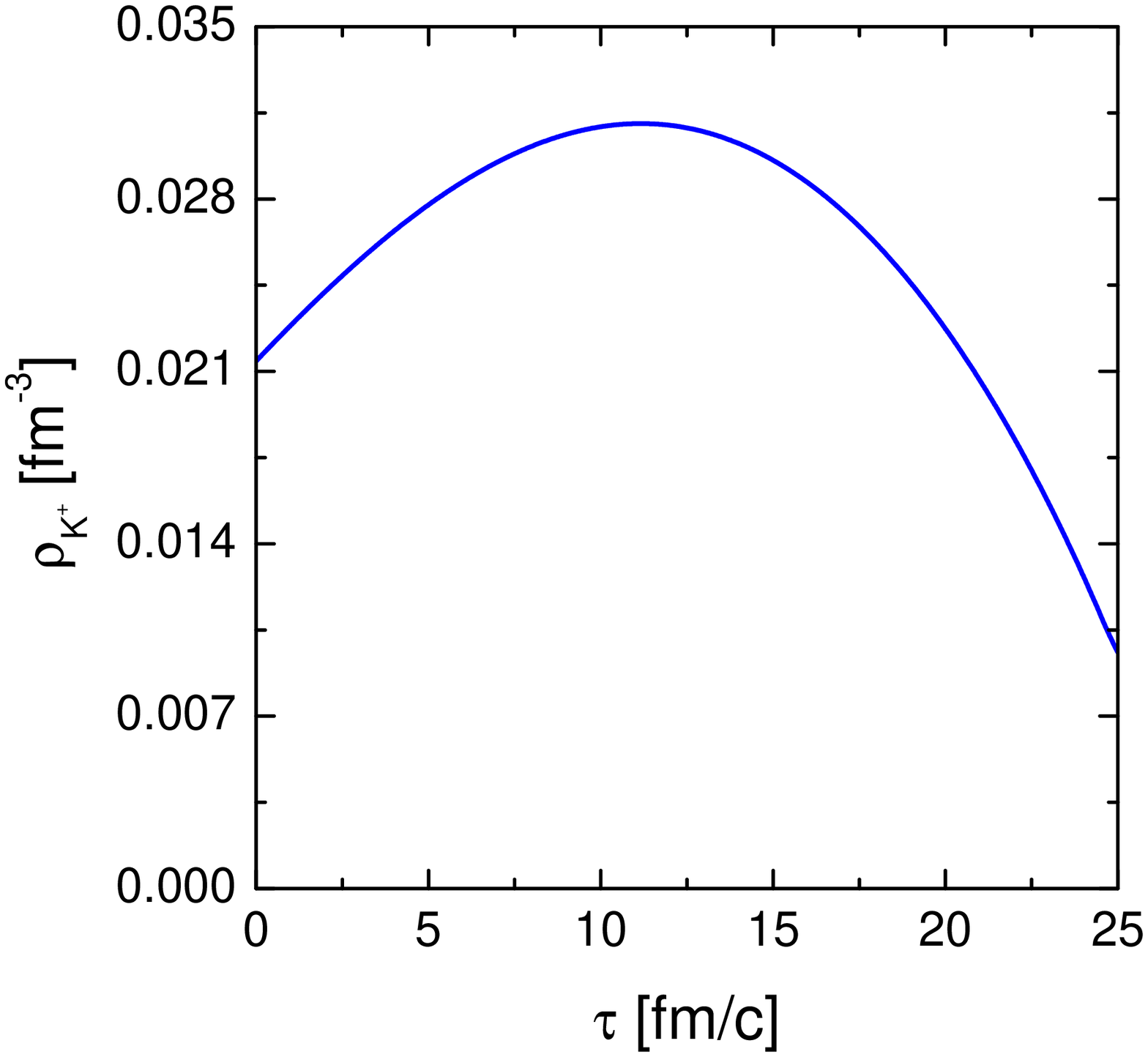}%
\includegraphics[width=.3\textwidth]{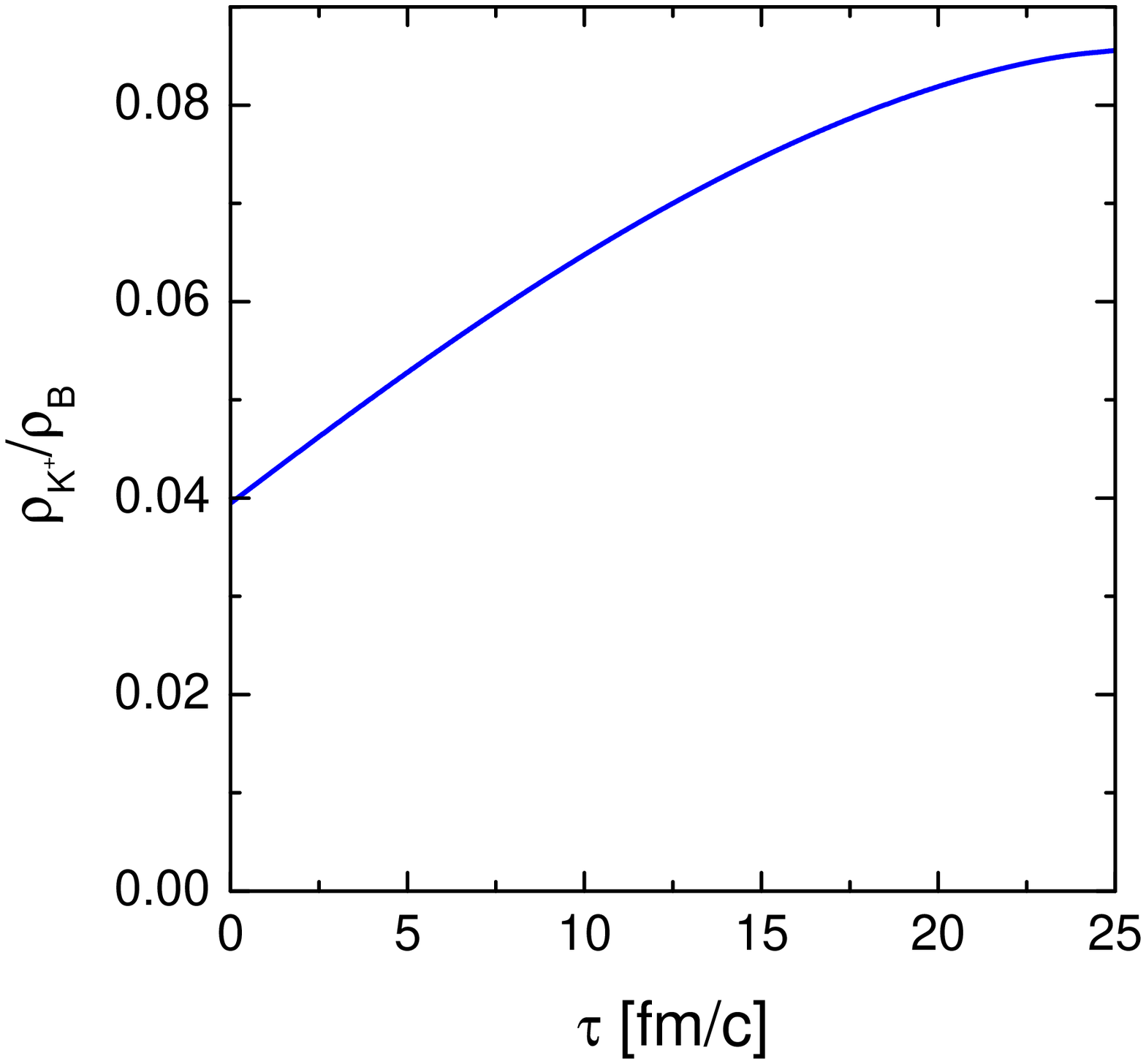}%
\includegraphics[width=.3\textwidth]{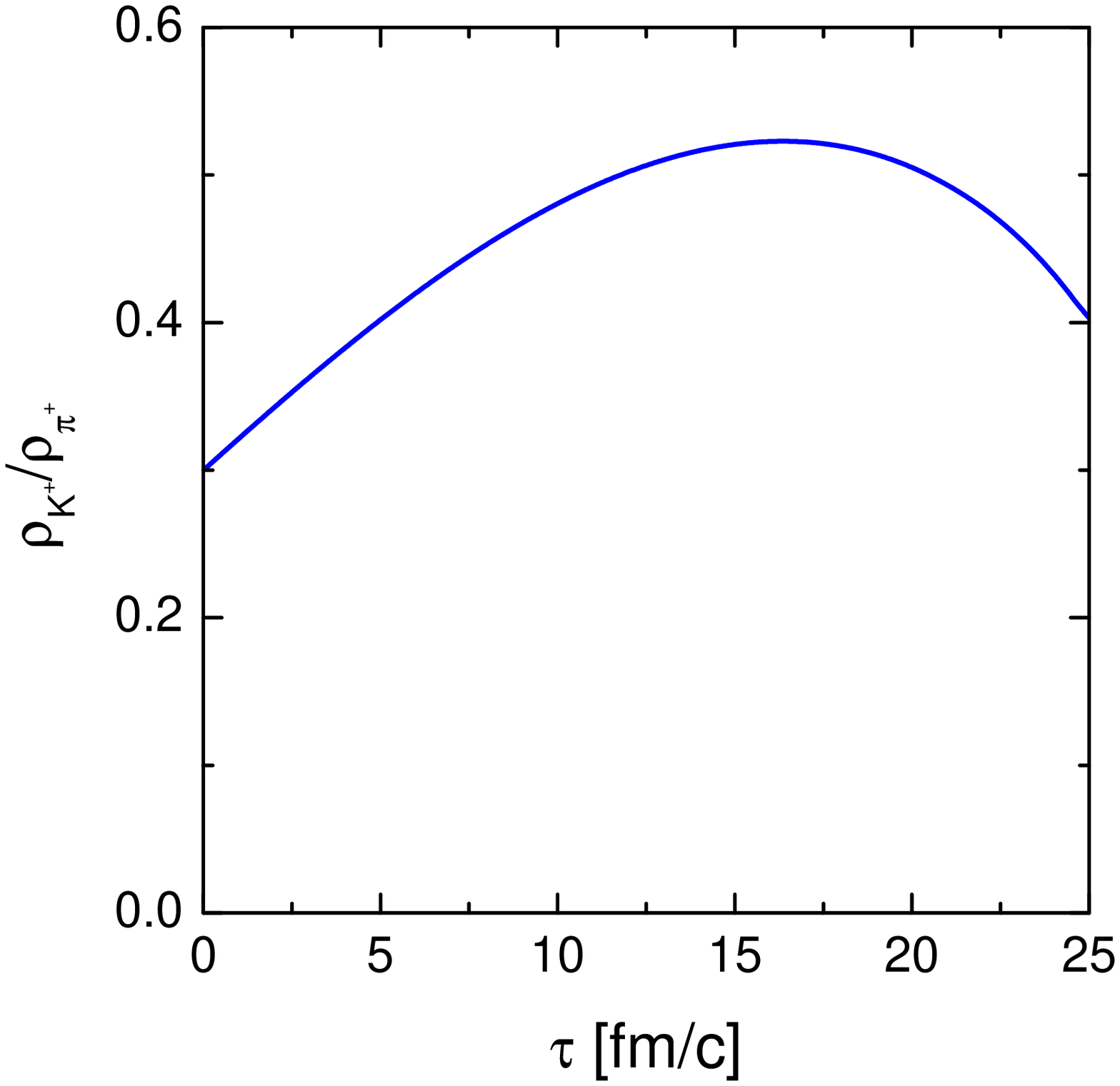}%
} \caption{The evolution of $K^+$ density (left), $K^+$ density
divided by baryon density as a density of conserved quantum charge
(middle), and  $K^+$ density divided by $\pi^+$ density.}
\label{f:kden}
\end{figure}
In Fig.~\ref{f:kden} we show the evolution of the density of
$K^+$. If only the density is plotted, we observe rise followed by
rapid decrease; the former is due to quick kaon production in a
slowly expanding fireball with strongly under-represented strange
particles, while the latter is mainly due to strong expansion and
corresponding decrease of all densities. In order to extract the
net effect of kaon production one can divide $K^+$ density by
baryon density which drops with time only due to expansion. The
ratio $\rho_K/\rho_B$ increases steadily, though the growth
becomes slower at later times when kaon abundance becomes larger.

Finally, we also show in Fig.~\ref{f:kden} the time-dependence of
the ratio of densities of positive kaons and pions. The final
value is about a factor of 2 larger than data value of 0.23, so it
may seem confusing if we say that this scenario reproduces data.
The point is that in the value which is compared with data we also
take into account feed-down from resonance decays, which is not
yet included in Fig.~\ref{f:kden}. The feed-down contributes more
to pion density in the denominator and brings the final value
close to measured data.

We want to close this section with a brief explanation of how the
final density of various strange species can be controlled. We
shall calculate ratios of multiplicities $K^+/\pi^+$, $K^-/\pi^-$,
and $\Lambda/\pi$. Density of kaons with $S>0$ is produced
kinetically and depends dominantly on the total lifespan and
slightly on temperature. Thus the lifespan is decisive for ratio
of $K^+/\pi^+$. Species with $S<0$ must balance strangeness such
that total $S$ of the system vanishes. Strange quarks are
distributed among $K^-$ and hyperons according to equilibrium
distribution. Hence, once we have the correct amount of $K^+$, the
final state temperature determines if we obtain correct results
also for $K^-$ and $\Lambda$'s.


\section{Results and discussion}

\begin{figure}[t]
\centerline{\includegraphics[width=.7\textwidth]{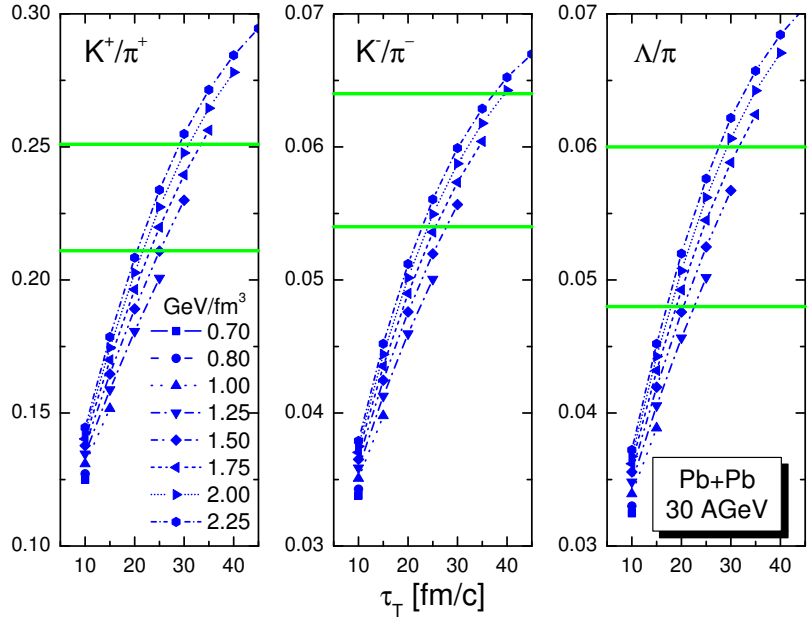}}
\caption{ Dependence of the resulting ratios of $K^+/\pi^+$
multiplicities (left), $K^-/\pi^-$ (middle), and $\Lambda/\pi$
(right) on total lifespan of the fireball calculated for Pb+Pb
collisions at projectile energy 30~$A$GeV. Different curves
correspond to different initial energy densities according to
legend.  \label{f:Rthirty}}
\end{figure}

\begin{figure}[h!]
\centerline{\includegraphics[width=.7\textwidth]{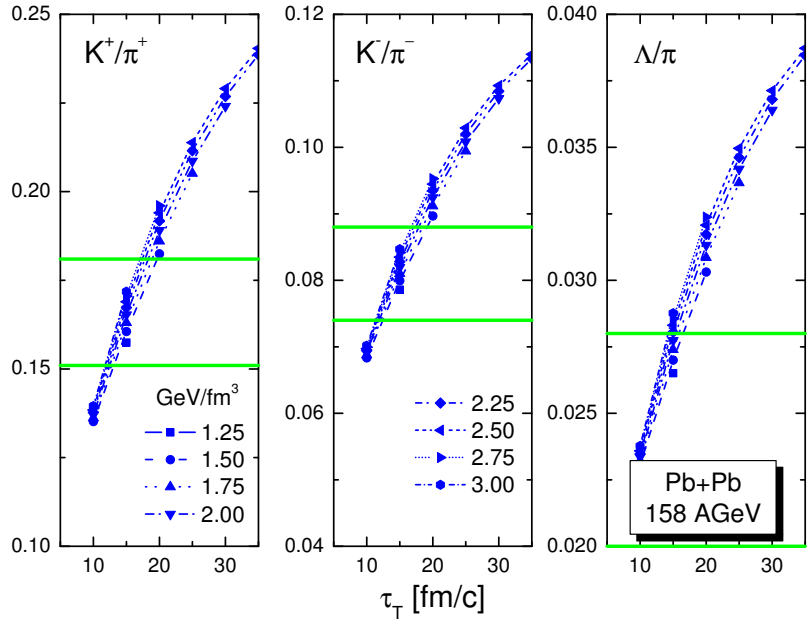}}
\caption{Same as Fig.~\protect\ref{f:Rthirty}, but for Pb+Pb
collisions at projectile energy 158~$A$GeV. \label{f:Rhfe}}
\end{figure}

\begin{figure}[t]
\centerline{\includegraphics[width=.7\textwidth]{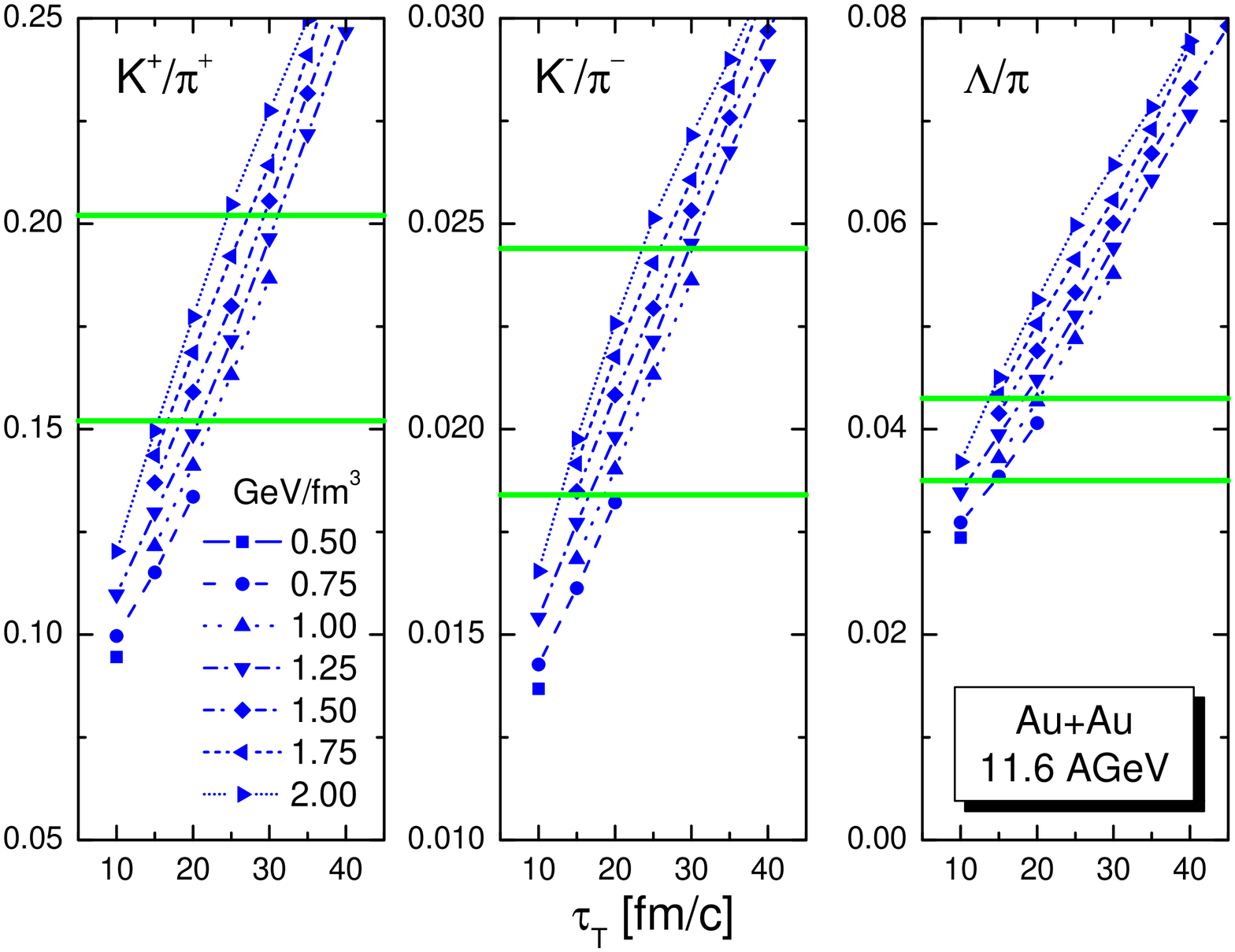}}
\caption{Same as Fig.~\protect\ref{f:Rthirty} but for Au+Au
collisions at projectile energy 11.6~$A$GeV.  \label{f:RAGS}}
\end{figure}

Typical results are presented in Figs.~\ref{f:Rthirty} (Pb+Pb at
projectile energy 30~$A$GeV) , \ref{f:Rhfe} (158~$A$GeV), and
\ref{f:RAGS} (Au+Au at 11.6~$A$GeV).
We observe that the ratios of strange particles to pions depend
strongly on the total lifespan of the fireball. One should not get
confused. What is plotted is dependence of the {\em final state}
ratio on {\em total lifespan}; it is {\em not} the evolution of
the ratios with time. In other words, every plotted point in the
figures corresponds to a different parametrisation of fireball
evolution.

Compared to the dependence on total lifespan, the dependence on
initial energy density appears less pronounced. It is more
important in collisions at lower energies which produce matter at
higher net baryon densities.

\begin{figure}[t]
\centerline{\includegraphics[width=.4\textwidth]{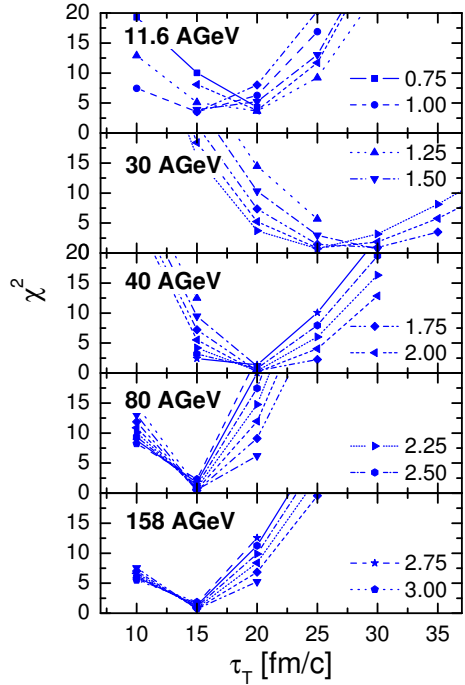}}
\caption{Values of $\chi^2$ determined from
Eq.~\protect\eqref{chid}. Different panels correspond to
collisions at different energy. $\chi^2$ is shown as a function of
total lifespan; different curves represent different initial
energy densities.} \label{f:chi}
\end{figure}

In order to reasonably compare our results with data, for every
triple of calculated ratios we construct a quantity $\chi^2$
which is determined as
\begin{equation}
\label{chid}
\chi^2 = \sum_{i=1}^{3} \frac{(d_i-t_i)^2}{e_i^2}
\end{equation}
where $t_i$, $d_i$, and $e_i$ stand for theoretical value, data
value, and its error, respectively. The sum goes over the three
density ratios calculated here. The values of $\chi^2$ calculated
according to this prescription are summarised in Fig.~\ref{f:chi}.
With this figure we want to test the hypothesis put forward in
Fig.~\ref{f:cartoon}: can a combination of increasing available
energy and {\em decreasing} total lifespan explain the observed
horn? From Fig.~\ref{f:chi} we see that the horny structure of
$K^+/\pi^+$ ratio is basically translated into a similar
dependence of the ``best'' total lifespan on collision energy. One
could speculate if this is really the case that the fireball lives
longest at lowest SPS energy. This could be reminiscent of an
evolution in the soft region of the equation of state. Still,
instead of making such a conclusion we want to note that the
quality of data does not allow to draw this kind of conclusions
very clearly.
In fact, one can still reproduce  data with rather acceptable
quality by scenarios whose lifespans do not increase with the
collision energy. We summarise  our choice of such scenarios in
Table~\ref{t:scen}.
\begin{table}[b]
\begin{center}
\begin{tabular}{c|ccccc} \hline\hline $E_{\rm beam}$ [$A$GeV]
& 11.6 & 30 & 40 & 80 & 158
\\ \hline
$\varepsilon_0$ [GeV/fm$^3$] & 1 & 1.5 & 2 & 2.25 & 2.75
\\
$\tau_T$ [fm/$c$] & 25 & 25 & 20 & 15 & 15
\\
\hline $T$ [MeV] & 118.1 & 139.0 & 147.6 & 153.7 & 157.8
\\
$T_f$ [MeV] & 114.7 & 134.1 & 143.3 & 149.3 & 153.6
\\ \hline\hline
\end{tabular}
\end{center}
\caption{%
Initial energy densities and total lifetimes from parameter sets
which were used in calculations leading to results shown in
Fig.~\protect\ref{f:comp}. In the lower portion of the table $T_f$
is the final state temperature obtained in our simulations and $T$
the temperature from the analysis of chemical freeze-out
\cite{becat}. } \label{t:scen}
\end{table}

\begin{figure}[h]
\centerline{\includegraphics[width=.7\textwidth]{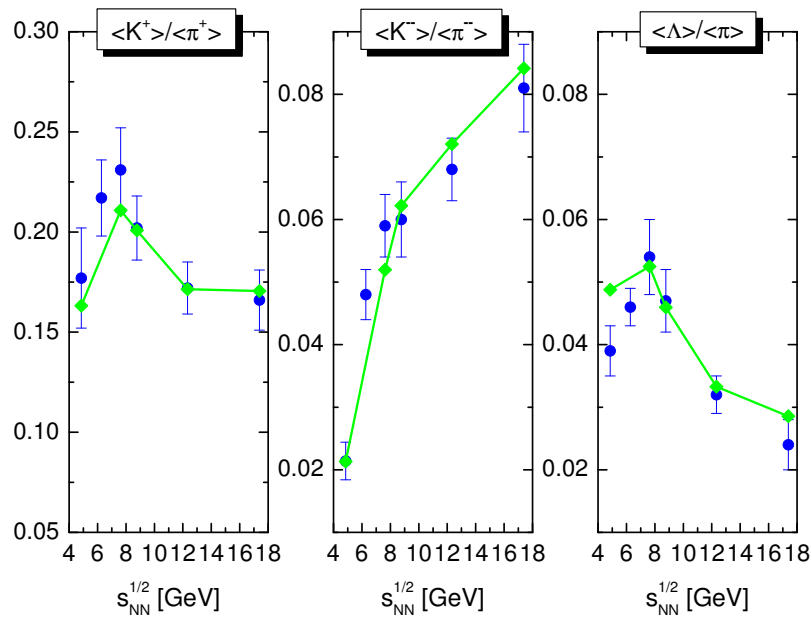}}
\caption{Comparison of data with selected results of our
simulations. Plotted are excitation functions of multiplicity
ratios of $K^+/\pi^+$ (left panel), $K^-/\pi^-$ (middle), and
$\Lambda/\pi$ (right). Data points are from \protect\cite{lung}. }
\label{f:comp}
\end{figure}

Finally, in Fig.~\ref{f:comp} we show the comparison of data to
results obtained with scenarios displayed in Table~\ref{t:scen}.
The data are reproduced rather well. Note that we did not perform
any calculations for the data point at projectile energy of
20~$A$GeV (data points at second lowest energies) as no tabulated
data were available to us. The slight failure of our model at
lowest collision energies is due to disagreement in the
$\Lambda/\pi$ ratio.

\section{Conclusions}

We managed---with a solely hadronic kinetic model---to reproduce
the famous ``horn'': excitation functions of multiplicity ratios
of $K^+/\pi^+$, $K^-/\pi^-$, and $\Lambda/\pi$. Therefore, merely
these data do not suffice for rejection of hadronic model of
fireball evolution and do not allow for convincing claims that
deconfinement sets in at collision energy where the horn appears.
In our framework, description of data requires decreasing total
lifetime as a function of collision energy at least for beam
energies above the horn.

Hence, in order to reject hadronic description comparisons with
other types of data must be undergone. The model---if
applicable---must be able to describe also single-particle spectra,
correlations, as well as total abundences. Dilepton spectra
appear as particularly interesting since dileptons are emitted
throughout the whole evolution of the fireball and thus are
sensitive to the total lifespan.

Finally, so far we put off the question, what microscopic
mechanism or equation of state leads to a total lifespan
decreasing with increasing collision energy. If our hadronic
kinetic model in which time evolution is only parametrised
survives all data tests, the quest for underlying microscopic
mechanism must be taken up.

\acknowledgments
BT thanks the organisers for invitation and kind hospitality in
Florence. He also thankfully acknowledges OZ Pr\'iroda for partial support
of the trip. The work presented here has been supported by a
Marie Curie Intra-European fellowship within the 6th European
Community Framework programme (BT) and by the US Department of Energy
under contract No.\ DE-FG02-87ER40328  (EEK).



\begin{thebibliography}{99}
\bibitem{roland}
G.\ Roland, these proceedings.

\bibitem{lung}
B. Lungwitz for the NA49 collaboration, AIP Conf.\ Proc.\
\textbf{828} (2006) 321.

\bibitem{smes} M. Ga\'zdzicki, M.I. Gorenstein, Acta Phys.
Pol. B. \textbf{30} (1999) 2705

\bibitem{kincal} J.K. Nayak, J.e. Alam, P. Roy, A.K.
Dutt-Mazumder, B. Mohanty, Acta Phys.\ Slov.\ \textbf{56} (2005)
27.

\bibitem{brat}
E.L.\ Bratkovskaya {\em et al.}, Phys.\ Rev.\ C \textbf{69} (2004) 054907.

\bibitem{wagner}
M.\ Wagner, A.B.\ Larionov, U.\ Mosel, Phys. Rev. C \textbf{71} (2005) 034910.

\bibitem{toneev}
Yu.B.\ Ivanov, V.N.\ Russkikh, V.D.\ Toneev,
Phys.\ Rev.\ C \textbf{73} (2006) 044904.

\bibitem{statmodel} J. Cleymans,  H. Oeschler, K.\ Redlich, S.\
Wheaton, Phys. Lett.\ B \textbf{615} (2005) 50.

\bibitem{becat} F. Becattini, M. Ga\'zdzicki, A. Ker\"anen, J.
Manninen, R.\ Stock, Phys.\ Rev.\ C \textbf{69} (2004) 024905.

\bibitem{rafel}
J. Rafelski, Phys.\ Rep.\ \textbf{88} (1982) 331.

\bibitem{bjork}
J.D.\ Bjorken, Phys.\ Rev. D \textbf{27} (1983) 140.

\bibitem{kora} P.\ Koch and J.\ Rafelski, Nucl.\ Phys.\ A \textbf{444}
(1985) 678.

\bibitem{komura} P.\ Koch, B.\ M\"uller, J.\ Rafelski, Phys.\ Rep.\
\textbf{142} (1986) 167.

\bibitem{kame} J.I.\ Kapusta and A.\ Mekjian, Phys.\ Rev.\ D \textbf{33}
(1986) 1304.

\bibitem{knoll} H.W.\ Barz, B.L.\ Friman, J.\ Knoll, H.\ Schulz,
Nucl.\ Phys.\ A \textbf{484} (1988) 661.

\bibitem{knoll2} H.W.\ Barz, B.L.\ Friman, J.\ Knoll, H.\ Schulz,
Nucl.\ Phys.\ A \textbf{519} (1990) 831.

\bibitem{brko} G.E.\ Brown, C.M.\ Ko, Z.G.\ Wu, L.H.\ Xia, Phys.\
Rev.\ C \textbf{43} (1991) 1881.

\bibitem{ktopi} B.\ Tom\'a\v{s}ik and E.E.\ Kolomeitsev, nucl-th/0512088.

\bibitem{dau} L.D.\ Landau, Izv.\ Akad. Nauk SSSR, Ser.\ Fiz., \textbf{17}
(1953) 51.

\bibitem{garo}
M. Ga\'zdzicki and D. R\"ohrich, Z.\ Phys.\ C \textbf{71} (1996)
55.


\end{thebibliography}
\end{document}